\documentclass{jaa}
\usepackage{natbib}
\usepackage{amsmath}
\newcommand{\mycomment}[1]{}

\usepackage{graphicx}

\begin{document}

\title{Exactly solved model of a one dimensional self gravitating system}


\author{Rajaram Nityananda
\textsuperscript{1,*}}
\affilOne{\textsuperscript{1}. International Centre for Theoretical Science, Tata Institute of Fundamental Research, Bengaluru 560089, INDIA\\}


\twocolumn[{

\maketitle

\corres{rajaram.nityananda@icts.res.in}

\msinfo{19 September 2025}{???}

\begin{abstract}
  
A model  one-dimensional self consistent steady state collisionless self-gravitating system in which all the particles have the same energy is presented.  This has the remarkable property that the position and velocity of the particles orbiting  in their own self consistent potential are given exactly, in terms of time,  by the truncations of sine and cosine functions  to the first two terms  in their  respective  Taylor series.  The potential and density also  have simple analytic expressions in terms of  time as parameter. It is not being claimed that  this system has any  direct astronomical application. However, it does motivate a conjecture about the behaviour of the density, potential, and orbits  near caustics in simulations of cold collisionless dark matter.  It is a rather surprising result which might interest practitioners of stellar dynamics and serve as an elementary  example in teaching the subject.  
\end{abstract}


}]


\doinum{12.3456/s78910-011-012-3}
\artcitid{\#\#\#\#}
\volnum{000}
\year{0000}
\pgrange{1--}
\setcounter{page}{1}
\lp{1}

\section{Introduction}
The isothermal    model for the density  distribution of stars perpendicular to a galactic disc is an example of a one dimensional  steady state self consistent self gravitating system (Spitzer 1942). Such a system can be constructed as follows. The distribution function in phase space is denoted by $f(x,v)$ and by Jeans theorem (Binney and Tremaine 1987 is the standard text), is taken to be a function of the conserved  energy per unit mass $E=v^2 + \phi(x)$, guaranteeing a steady state.  This shows  that the real   space matter density, \(\rho(x)\) is a  function of the gravitational potential \(\phi(x)\), and solves the one dimensional Poisson equation, 
\begin{equation}
   \frac{\mathrm{d}^2 \phi(x) }{\mathrm{dx}^2}  =4 \pi G \rho(\phi) 
\end{equation}
This formulation ignores the discreteness of the system and is valid in the limit of an infinite number of particles (sheets in one dimension) each  with vanishing   surface density,  keeping the overall volume  density finite. Such a system is   described as collisionless,   since the only interaction between the sheets is via their own mean potential. 
In the Spitzer model,  \( \rho(\phi)=A \exp(-\beta \phi) \). After multiplying both sides by \( \mathrm{d} \phi(x) /{\mathrm{d}x }  \) and integrating twice, one obtains a solution of the form  \(\rho(x) \propto {\mathrm{sech}^2(Cx)} \), and  a phase space distribution function \(f(x,v)=f_0 \exp (-\beta( v^2/2 +\phi(x) )\)
This describes a stellar system  with a gaussian distribution of velocities  which is independent of the  position \(x\) . 

There is an extensive  literature, both numerical and analytical, on  such systems. This includes  dynamics, steady states, and approach to 'thermal' equilibrium (the Spitzer state) induced by collisions when the particle number  $N$ is finite. There is an  excellent recent reference which also   points to the earlier literature (Sousa and Rocha Filho, 2023).  We are concerned in this brief note with a purely analytical construction of  a very specific steady state system, for its own sake, not claiming a close connection with this literature, or real galactic discs. At the end, we note the instability of this solution and demonstrate  its  relaxation to a more disordered state. using a numerical simulation.
\section{The constant energy model}
The system which we consider is in some sense the opposite of the Spitzer model.  The phase space distribution function  is a delta function of the total single particle energy per unit mass,  \(E\) This becomes the real space density after integrating over $v$, so that its dimensions are $ML^{-4}T$.  $f_0$ has dimensions $ML^{-2}T^{-1}$ allowing for the delta function which follows it. 
 Integrating over \(v\) to obtain the real space mass  density, we get 
\begin{equation}    
 \rho(x) =2 f_0~ / ~|v(x)|=2 f_0 /\sqrt{2(E-\phi(x))} 
 \end{equation} 
 
 This  has the physical interpretation that the time spent by any  particle  in a small region  is inversely proportional to its  speed, which  is a unique function of the position thanks to the fixed total energy. 
 We choose units so that  \(G=1/4\pi\),  $f_0=1/2$ and $E=1/2$.    Note that the powers of  mass, length,  and time, occurring in these three quantities   are    $ [G]=[-1,3,-2]$, \hspace{0.1 cm}$[f_0] =[1,-2,-1] $,\hspace{0.1 cm}$ E=[0,2,-2] $.  The determinant formed from these three vectors is nonvanishing. Therefore  the choices made  lead to  unique units of $M,L,T$ . Upper case letters are used to denote the position $X$, the velocity $V$, the potential $\Phi$ and the acceleration $A =-d\Phi/dX $ in these units. The right hand side of the Poisson equation is now simply the reciprocal of the speed. 
 
 \subsection{The parametric solution } The main point of this note is that there is a remarkably simple parametric form for the potential and the orbit.    The Poisson equation, in our units, reads
 
\begin{equation}
  \frac{ d^2 \Phi}{dX^2} \equiv  -\frac {dA}{dX }=1/V
\end{equation}  
This leads to  
\begin{equation}
V\frac {dA}{dX }=\frac{dA}{dT}=-1    
\end{equation}
The origin of time is chosen  when the particle passes the midplane, moving to positive $X$. At this time,  the force and hence  acceleration vanish. The potential can be chosen to be zero at $X=0$. Using these conditions, we can integrate the previous equation to get the acceleration, the velocity,  the position, and the potential  in terms of time.
\begin{equation}
A=-T
\end{equation}
\begin{equation}
V=1-T^2/2
\end{equation}
\begin{equation}
X=T-T^3/6
\end{equation}
\begin{equation}
\Phi=E-V^2/2= \frac{1}{2}-\frac{1}{2} (1-T^2/2)^2 = \frac{1}{2}T^2- \frac{1}{8}T^4
\end{equation}

 Note that the forms for position and velocity are two-term truncated Taylor series of the  sine and cosine which would have appeared for a harmonic oscillator. This oscillator is not harmonic. As written, the motion is not even periodic, but we regard  these expressions as  representing  just one half cycle of the oscillation, during which the velocity is positive, and hence to be used in the range 
\(-\sqrt 2 <T <\sqrt 2\).  In the adjacent half cycle, \( \sqrt 2 <T< 3 \sqrt 2 \), centred on $T=2\sqrt 2$, the velocity becomes
\(V=-1`+(T-2\sqrt(2))^2/2\) and the position is given by \( X= -(T-2\sqrt 2) +(T-2\sqrt(2))^3/6 \). Beyond this range, they are periodically repeated with period $4 \sqrt 2$ Note that at the points $T= -\sqrt 2 +  2 \sqrt 2n$, with $n$ an integer, the second derivative of the velocity changes sign from $-1$ to $+1$
Figure 1 shows a quarter cycle of the motion  for the particles of the model, as well as for  a reference harmonic oscillator. ( denoted by 'ho' in the figure).  This is  chosen to have the same period and energy,   The potential of this oscillator can be calculated from the period as  $ \phi _{ho} =  (1/2) \omega^2 x^2=(1/2)(2\pi/4 \sqrt(2))^2 x^2= (1/2)(\pi^2 /8) x^2$. This choice of reference  is not unique, one could have chosen to match the amplitude in the co-ordinate, fer example. 
\begin{figure}
    \centering
    \includegraphics[width=1.0\linewidth]{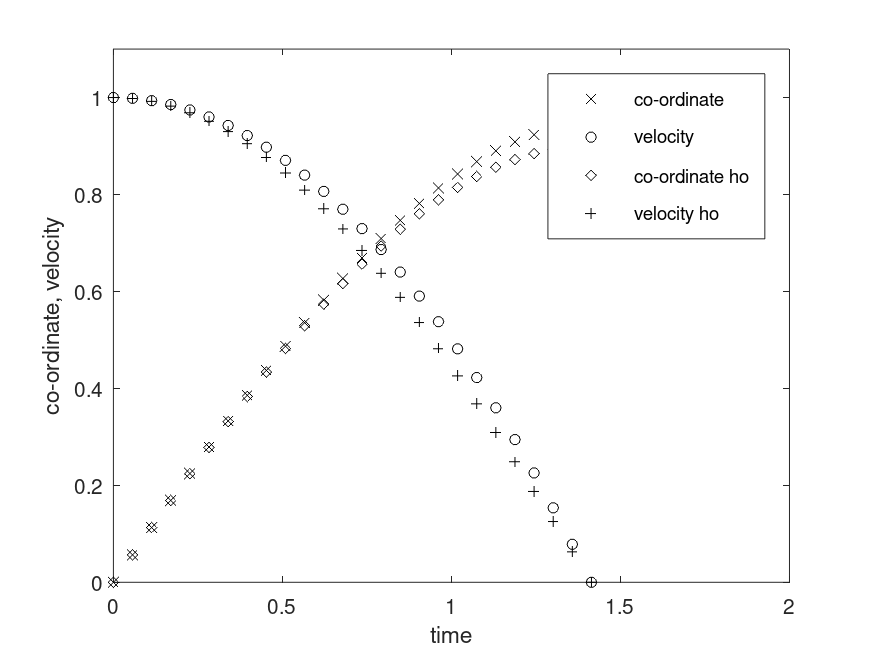}
    \caption{The position and velocity vs. time  are plotted for the one dimensional motion of the particles in the model, over a quarter  cycle, For comparison, the same quantities for the reference harmomic oscillator are plotted as well}
    \label{fig:xvt}
\end{figure}

 It might have seemed more  natural to set the system size $l$,  half period of oscillation $t$, and total mass per unit area $\sigma$ to unity.  The price of the simple parametric form is that our units are related to $l$,$t$,and $\sigma$ by  numerical factors.  The range of $X$ is $ 4 \sqrt(2)/3  $,  so our unit of length is clearly the system size $l$  divided by $4\sqrt(2) /3 \approx 1.89$. The unit of time, similarly,  is the half period of oscillation  $t$ divided by $2 \sqrt (2) \approx 2.83$. The  surface density  $\Sigma$ is obtained by integrating the real space density $1/V$ with respect to $X$, giving $\Sigma=2 \sqrt(2)  $. So the unit of surface density is the system surface density  $\sigma$ divided by $2 \sqrt(2) \approx 2.83$

 \subsection{Discussion of the solution}

\begin{figure}[t]
    \centering
    \includegraphics[width=1.0\linewidth]{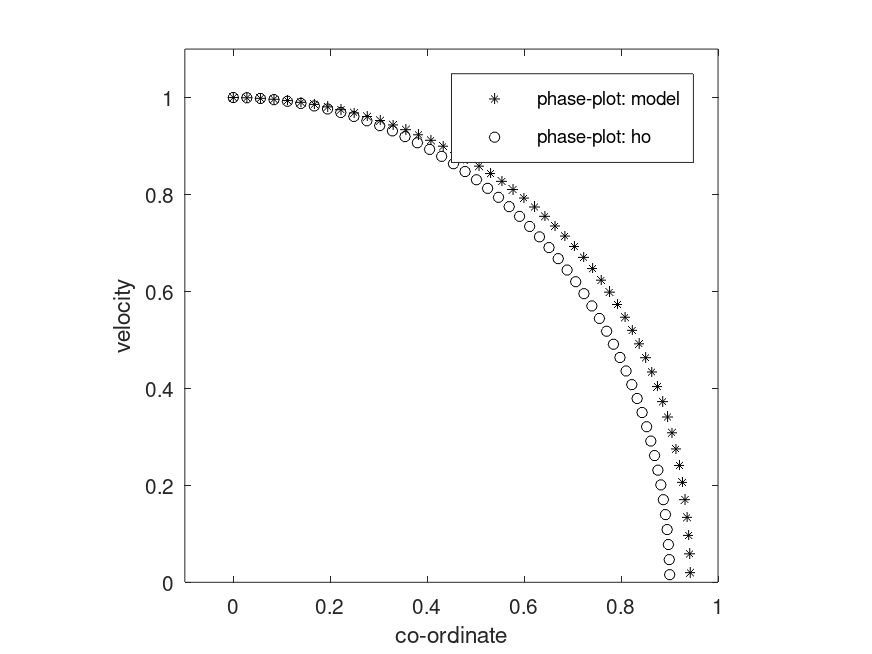}
    \caption{Phase trajectory of the motion in the X-V plane for a quarter cycle. The trajectory  for the reference harmonic oscillator is shown for comparison}
    \label{fig:phaseplot}
\end{figure}
\begin{figure}[h!]
    \centering
    \includegraphics[width=1.0\linewidth]{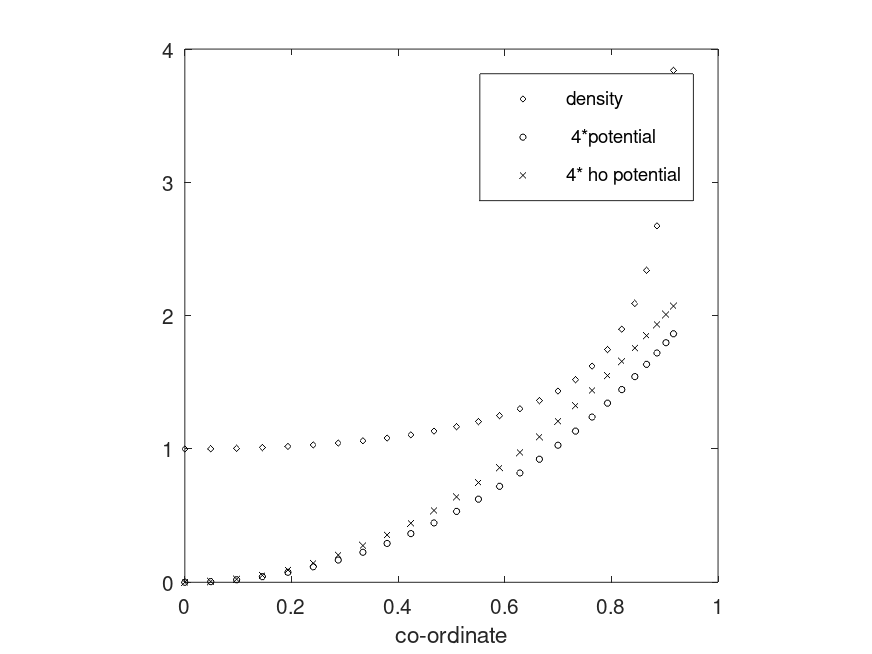}
    \caption{The self consistent density and potential as functions of position. The potential has been multiplied by 4, its actual value at the ends is 0.5. The reference harmonic oscillator potential also scaled by 4, is plotted as well The density plot has been truncated, since  it rises to infinity at the two extreme values of x}
    \label{fig:densitypotential}
\end{figure}
To the  eye, the phase trajectories as well as the time dependence look remarkably like those of a simple harmonic oscillator, and so does the potential. It is the self consistent density which reveals the difference.  It  is not constant, as would be needed to create a harmonic oscillator potential, but has a sharp rise at the turning points. 
The singularity of the density in Figure 3.  is  brought out  by Taylor expansion of $X$ and $V$ about $T=\sqrt 2$, the turning point of the oscillator. The position has a maximum there, while the velocity goes to zero linearly, so it is clear that $V$ scales as $ (2\sqrt2/3-X)^{1/2}$, and the density $1/V$  hence has an  integrable, inverse square root singularity at the turning point. This kind of behaviour is characteristic of a 'fold catastrophe' and is also seen in caustics in optics e.g of gravitational lenses, and in simulations of cold dark matter with smooth initial conditions (Shandarin and  Zeldovich 1989) .  A  question then arises. How does   the singularity in the density affect the motion of the dark matter particles in such cold simulations?  The question  is rather   academic, since only the second derivative of the potential has this integrable singularity. This question can be answered in this model - the  velocity and acceleration of the particles are continuous as they  touch  the caustic and return, but the time derivative of the acceleration has a simple discontinuity - a step  in the jerk so to speak. Since this is local phenomenon near the turning point in phase space, it is natural to conjecture that it will hold in more general situations.  Of course,  in any more realistic situation with velocity dispersion,  one expects  that the caustic,   will  be smoothed out, and with it this higher order discontinuity  in the trajectory.
\subsection{Instability: some numerics}
"Cold" models in stellar dynamics tend to be unstable.  For  example, there is a well known criterion  predicting a bar  instability of a  model galaxy  if the kinetic energy in ordered rotational motion  is more than a fraction, about seven per cent,  of the total kinetic energy ( Ostriker and Peebles 1973 )  The model  of this note is cold in this sense,  with all  the kinetic energy in an ordered form,  concentrated in two opposing  streams at any given point.  

A simple simulation with 30 sheets, attracting each other with a constant force, as appropriate to one dimension, was  carried out, essentially to machine accuracy.  The force  on a given sheet is a constant, and  is proportional to the difference between the numbers of sheets on its two sides, until that number changes. One therefore evolves the whole system exactly till the first  pair of sheets  cross,  interchanges their order, and evolves till the next crossing, etc.  Figure 4 shows the initial phase space  distribution  chosen to approximate the model of this paper. For a few crossing times, the distribution stays fixed as the particles move along the same fixed orbit, indicating  a stationary situation.  Given the discretization,  this is not strictly true  and the deviations from the initial distribution grow.   Figure 5 shows the phase space distribution after 60 times the initial half-period of 2.8.  This  is a  much 'hotter' distribution  with a range of particle energies.    To test that it is a true dynamical effect rather than  a numerical artifact,  the velocities were reversed in the final  state and the initial state was recovered satisfactorily. 

\begin{figure}[h!]
    
    \includegraphics[width=0.4\textwidth]{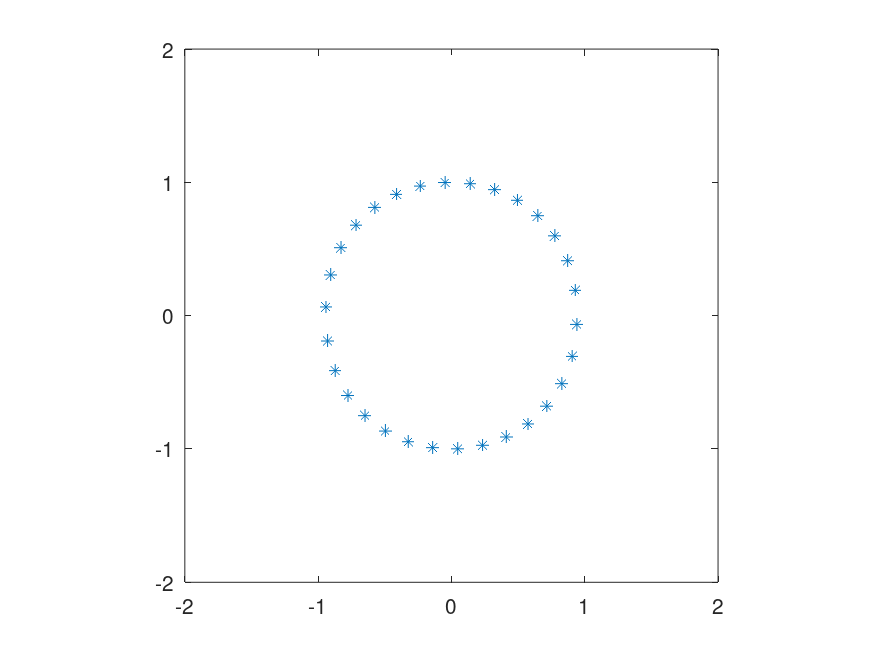}      
    \caption{ The initial phase space distribution of the model discretized using     30 sheets}
    
\end{figure}
\begin{figure}[h!]
    
    \includegraphics[width=0.4\textwidth]{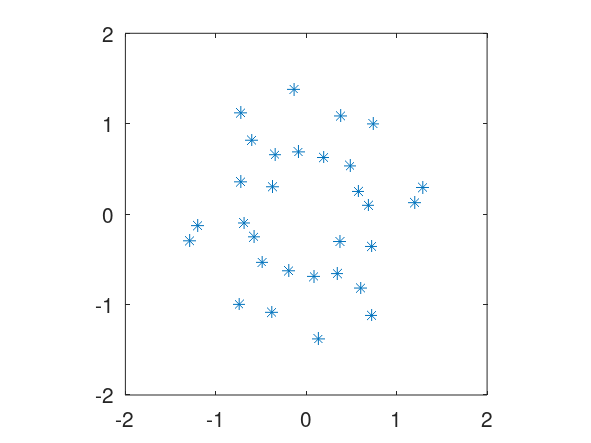}      
    \caption{ The phase space distribution afer 60 crossing  times }
    
\end{figure}

\section{Conclusion }
A one dimensional steady state self consistent  self gravitating system is constructed.  The " particles" (sheets interacting with a $|x|$ potential ) all move  on a single orbit with a fixed energy.  When time along the orbit is chosen as a parameter, all the quantities take a very simple algebraic form, (equations 6 to 9 ) which I have not found in the literature. This model is unstable as expected.  It  gives a pointer to the weakly singular  behaviour of the orbits, density, and potential near the fold caustics which occur naturally in simulations of cold collisionless dark matter with  sufficiently smooth initial conditions. The feature  that the time derivative of the acceleration has a finite discontinuity as the orbit touches the caustic, is conjectured to be general.

\section*{Acknowledgement}
I thank two anonymous reviewers for many  suggestions which greatly  improved  the presentation.
Vanchana Rathore,  a  summer project student  from Delhi University, at the National Centre for Radio Astrophyics in Pune,  worked on this problem and showed me one could make  significant  analytical progress, leading  to the results presented here.   Vanchana  would naturally  have been the first author of this note, had I been able to locate her.  I acknowledge her primary role in this work, and  hope that she gets to  see it.  I have  shown this model  to  many people in the field  over two decades, and they did not point me to a prior reference. 

\vspace{-1em}








\end{document}